\documentclass[twocolumn,preprintnumbers,elsart]{revtex4-1}
\usepackage{makeidx}
\usepackage{amssymb}
\usepackage{amsmath}
\usepackage{graphicx}
\usepackage{dcolumn}
\usepackage[center]{subfigure}
\usepackage{float}
\usepackage{color}

\begin{document}

\title{Josephson oscillations of chirality and identity in two-dimensional
solitons in spin-orbit-coupled condensates}
\author{Zhaopin Chen$^{1}$, Yongyao Li$^{2,1}$}
\author{ Boris A. Malomed$^{1,3}$}
\address{$^{1}$Department of Physical Electronics, School of Electrical Engineering,
Faculty of Engineering, Tel Aviv University, Tel Aviv 69978, Israel\\
$^{2}$School of Physics and Optoelectronic Engineering, Foshan University,
Foshan 52800, China\\
$^{3}$Instituto de Alta Investigaci\'{o}n, Universidad de Tarapac\'{a}, Casilla 7D,
Arica, Chile}

\begin{abstract}
We investigate dynamics of 2D chiral solitons of semi-vortex (SV) and
mixed-mode (MM) types in spin-orbit-coupled Bose-Einstein condensates with
the Manakov nonlinearity, loaded in a dual-core (double-layer) trap. The
system supports two novel manifestations of Josephson phenomenology: one in
the form of persistent oscillations between SVs or MMs with opposite
chiralities in the two cores, and another one demonstrating robust periodic
switching (\textit{identity oscillations}) between SV in one core and MM in
the other, provided that the strength of the inter-core coupling exceeds a
threshold value. Below the threshold, the system creates composite states,
which are asymmetric with respect to the two cores, or collapses. Robustness
of the chirality and identity oscillations against deviations from the
Manakov nonlinearity is investigated too. These dynamical regimes are
possible only in the nonlinear system. In the linear one, exact stationary
and dynamical solutions for SVs and MMs of the Bessel type are found. They
sustain Josephson self-oscillations in different modes, with no
interconversion between them.
\end{abstract}

\maketitle



\section{Introduction}

Josephson oscillations, induced by tunneling of wave functions between
weakly coupled cores, is a ubiquitous effect in macroscopic quantum systems
\cite{book}. It has been predicted and observed in superconductors separated
by a thin insulating layer \cite{josephson1962}-\cite{super}, superfluid $%
^{3}$He \cite{davis2002superfluid}, atomic Bose-Einstein condensates (BECs)
\cite{Milburn}-\cite{Veronica}, optical couplers \cite{asymm1}-\cite{Ignac},
and exciton-polariton waveguides \cite{exciton}. Among other applications,
the Josephson effect in superconductors may be used for design of qubits
\cite{Levitov,Shnirman}. Josephson oscillations of angular momentum between
annular BECs was investigated too \cite{lesanovsky2007spontaneous}-\cite%
{Yakimenko3}. In these contexts, chirality of the wave functions may play an
important role \cite{lesanovsky2007spontaneous}, \cite{Averin}-\cite{Law}.

In its linear form, the Josephson effect only gives rise to plasma waves in
superconducting junctions. The most significant modes in the junctions are
topological solitons (\textit{fluxons}) supported by the nonlinear Josephson
relation between the tunneling current and phase difference across the
junction. Both classical \cite{Fulton}-\cite{Ustinov} and quantum \cite%
{Walraff} fluxons, as well as \textit{superfluxons}, i.e., kink excitations
in fluxon chains \cite{super}, were studied theoretically and experimentally.

New possibilities for realization of dynamical effects in BEC are offered by
binary (pseudo-spinor) condensates with spin-orbit coupling (SOC) between
their components \cite{SOC1}-\cite{SOC4}. In particular, SOC affects the
Josephson tunneling between annular condensates \cite{WMLiu}. Stable 2D
solitons, which include vortex components, were predicted in the SOC system
of the Rashba type \cite{Rashba} with attractive nonlinear interactions \cite%
{Sakaguchi_2014}. In terms of optics, these interactions are represented by
self- and cross-phase-modulation (SPM and XPM) terms in the underlying
system of coupled Gross-Pitaevskii equations. While in the absence of SOC,
attractive cubic terms in 2D equations always lead to collapse \cite{Fibich}
(i.e., the system does not have a ground state), the interplay of the
attractive SPM and XPM terms with SOC gives rise to ground states, in the
form of solitons of the \textit{semi-vortex} (SV, alias half-vortex \cite%
{half}) or \textit{mixed-mode} (MM) types, provided that, respectively, the
SPM nonlinearity is stronger or weaker than its XPM counterpart. SVs are
composed of a vortex in one component and zero-vorticity soliton in the
other, while MMs combine vortical and zero-vorticity terms in both
components. In the experiment, such states can be created by using helical
laser beams that resonantly couple to one component only, and thus transfer
the angular momentum onto that component, without exciting the other one
\cite{transfer1,transfer2,transfer3}. Two or several solitons in the layers'
plane can be created too, by means of an initially applied cellular in-plane
potential, which cuts the condensate into lateral fragments, and is lifted
afterwards \cite{Randy}.

If the SV soliton is the ground state, the MM is an unstable excited one,
and vice versa. Similar results were reported \cite{Sherman} for more
general SOC systems, of the Rashba-Dresselhaus \cite{Dressel} type.
Two-dimensional SV solitons exist in the form of two different \textit{%
chiral isomers (}right- and left-handed ones), with vorticity sets in their
components being, respectively, $\left( S_{+},S_{-}\right) =(0,1)$ or $%
(-1,0) $ (mirror images of each other). Similarly, there exist two different
chiral forms of MMs, which are introduced below in Eq. (\ref{MMansatz}).

In the experimental realization of SOC in binary BEC, two components are
represented by different hyperfine states of the same atom, with nearly
equal strengths of the SPM and XPM\ interactions, suggesting one to consider
the \textit{Manakov nonlinearity }\cite{Manakov}, with equal SPM\ and XPM
coefficients. In such a case, the system is invariant with respect to
rotation of the pseudo-spinor wave function in the plane of its two
components. The 2D system combining SOC and the Manakov nonlinearity gives
rise to an additional soliton family, which embeds the SV and MM solitons
into a continuous set of intermediate states \cite{Sakaguchi_2014}. The
family is a degenerate one, in the sense that all solitons with a fixed
total norm share common values of the energy and chemical potential. The
family is dynamically stable against small perturbations, but structurally
unstable, as a deviation from the SPM = XPM condition breaks the
intermediate states, keeping only the SV and MM solitons as robust modes.

The subject of this work is a junction formed by a pair of Josephson-coupled
2D layers (\textit{cores}), each one carrying spin-orbit-coupled BEC. In the
experiment, a two-layer setup can be realized by loading the condensate in
two adjacent valleys of a deep 1D optical lattice, which illuminates the
setting in the perpendicular direction (see, e.g., Ref. \cite{layers}). In a
general form, a dual-core 2D BEC system was introduced in Ref. \cite{Arik}.
In that work, 2D\ solitons, built of components in the two layers, were
stabilized by a spatially-periodic in-plane potential, while SOC was not
considered. The problem addressed in Ref. \cite{Arik} was spontaneous
breaking of the inter-core symmetry in the solitons, but not Josephson
oscillations. Here, we consider oscillations between Josephson-coupled 2D
SOC condensates, of both SV and MM types, in the dual-core system. In its
linear version, Bessel-shaped exact solutions are found in an analytical
form, for stationary states and Josephson oscillatory ones alike. These
solutions demonstrate solely intrinsic oscillations in the two components of
the condensate, which do not mix states of the SV and MM types, nor
different isomers in each type. In the full nonlinear system, solutions for
self-trapped states are obtained in a numerical form. They readily
demonstrate robust periodic oscillations between components of the 2D
solitons with \emph{opposite chiralities} in the two layers. Furthermore,
Josephson-type \emph{identity oscillations}, i.e., periodic mutual
interconversion between solitons' components of the SV and MM\ types in the
two layers, are reported too.

The rest of the paper is structured as follows. The model is formulated in
Section II, in which basic types of the considered states are introduced
too, including new solutions for the Bessel-shaped SV and MM\ modes in the
linear system. Results for the Josephson oscillations are collected in
Section III. It includes exact solutions obtained in the linear system, and
a summary of the systematic numerical analysis of Josephson oscillations in
2D solitons of different types in the full nonlinear system. The numerical
results make it possible to identify robust regimes of oscillations of the
chirality (right $\rightleftarrows $ left) and identity (SV $%
\rightleftarrows $ MM) between the coupled layers. The paper is summarized
in Section IV, where we also give estimates of the predicted effects in
physical units, and discuss directions for the further work; one of them may
be the use of beyond-mean-field effects \cite{Petrov1} for the stabilization
of the 2D\ system against the critical collapse \cite{Fibich}, when its norm
exceeds the respective threshold value.

\section{The model: coupled Gross-Pitaevskii equations}

\subsection{The single-layer spin-orbit-coupled (SOC)\ system}

For the single effectively two-dimensional layer, the system of coupled
Gross-Pitaevskii equations for components $\phi _{\pm }$ of the
pseudo-spinor wave function, coupled by the spin-orbit interaction of the
Rashba type, is written as \cite{Sakaguchi_2014,Sherman}
\begin{gather}
i\partial _{t}\phi _{+}=-\left[ \frac{1}{2}\nabla ^{2}+\left( |\phi
_{+}|^{2}+\gamma |\phi _{-}|^{2}\right) \right] \phi _{+}  \notag \\
+\left( \partial _{x}-i\partial _{y}\right) \phi _{-},  \label{phi+}
\end{gather}%
\begin{gather}
i\partial _{t}\phi _{-}=-\left[ \frac{1}{2}\nabla ^{2}+\left( |\phi
_{-}|^{2}+\gamma |\phi _{+}|^{2}\right) \right] \phi _{-}  \notag \\
-\left( \partial _{x}+i\partial _{y}\right) \phi _{+},  \label{phi-}
\end{gather}%
where $t$ and $x$, $y$ are scaled time and coordinates, the SPM and SOC
coefficients (the latter one is the coefficient in front of the first-order
spatial derivatives) are set to be $1$ by scaling, and $\gamma $ is the
relative XPM\ strength. We will chiefly address the Manakov nonlinearity,
with $\gamma =1$, which is close to the experimentally relevant situation
\cite{SOC1}-\cite{SOC4}; effects of deviation from the Manakov's case are
considered too. Equations (\ref{phi+}) and (\ref{phi-}) do not include an
external trapping potential, which is necessarily present in the experiment,
but its effect on stable solitons is negligible \cite{Sakaguchi_2014}.

Stationary self-trapped states of the pseudo-spinor condensate (solitons)
with chemical potential $\mu <0$ are looked for as
\begin{equation}
\phi _{\pm }(x,y,t)=\Phi _{\pm }(x,y)e^{-i\mu t},  \label{phiPhi}
\end{equation}%
with the 2D norm
\begin{equation}
N_{\phi }=N_{_{+}}+N_{_{-}}\equiv \int \int \left[ |\Phi _{+}\left(
x,y\right) |^{2}+|\Phi _{-}\left( x,y\right) |^{2}\right] dxdy.  \label{N}
\end{equation}

\begin{figure}[tbp]
\centering{\label{fig1a} \includegraphics[scale=0.24]{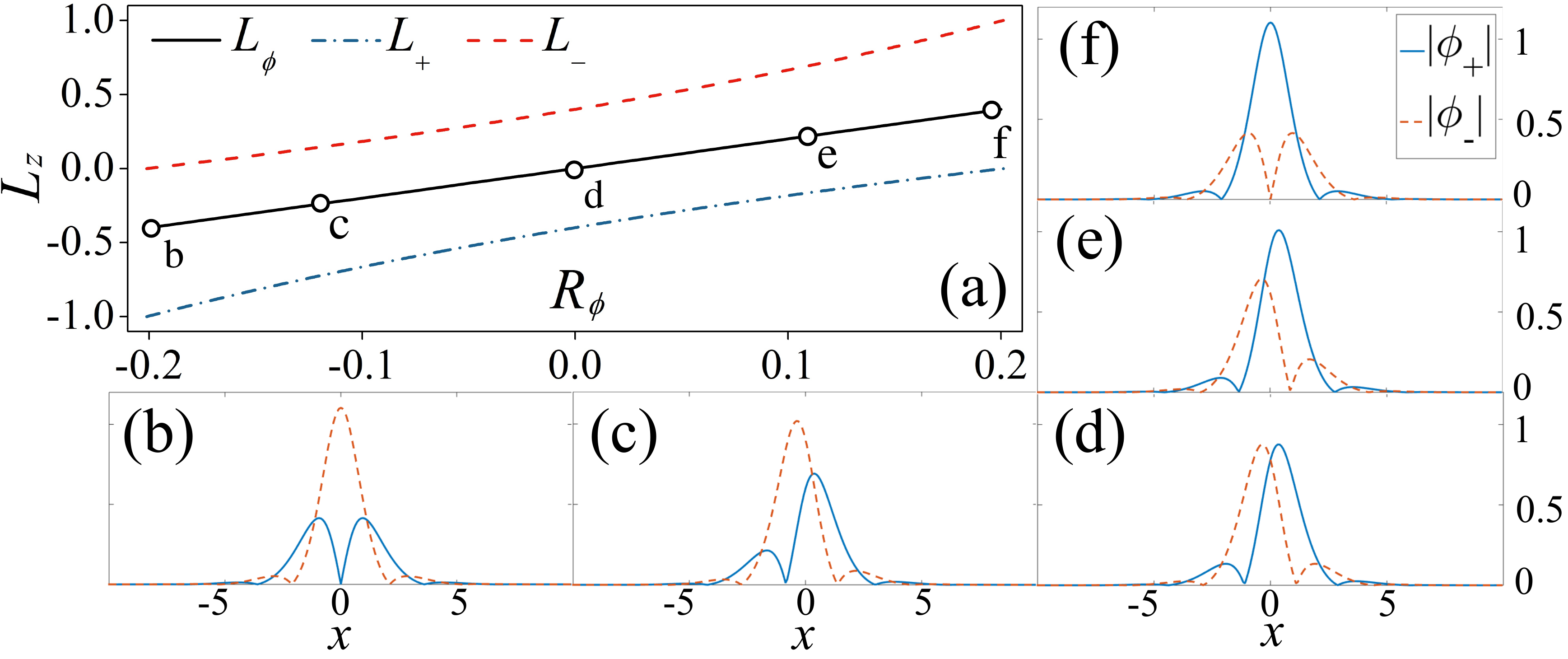}}
\caption{(Color online) (a) Angular momenta per particle ($L_{z}=L_{\protect%
\phi }$ or $L_{\pm }$, as indicated in the panel), which are defined by Eqs.
(\protect\ref{L}) and (\protect\ref{LL}), are shown as functions of norm
ratio $R_{\protect\phi }$ [see Eq. (\protect\ref{R})], for the continuous
family of stable SOC solitons, numerically generated by input (\protect\ref%
{M}), with fixed norm $N_{\protect\phi }=4$, in the single-core system.
Linear dependence $L_{\protect\phi }(R_{\protect\phi })$ follows Eq. (%
\protect\ref{linear}). The energy and chemical potential of all solutions
belonging to the degenerate family are $E_{\protect\phi }=-2.4$ and $\protect%
\mu =-0.87$. Cross-sections of two components of the 2D solitons, marked in
(a), are displayed for the following values of integral quantities (\protect
\ref{R}) and (\protect\ref{L}), (\protect\ref{LL}): $R_{\protect\phi }=-0.2$%
, $L_{\protect\phi }=-0.4$ (b); $R_{\protect\phi }=-0.12$, $L_{\protect\phi %
}=-0.24$ (c); $R_{\protect\phi }=0$, $L_{\protect\phi }=0$ (d); $R_{\protect%
\phi }=0.11$, $L_{\protect\phi }=0.22$ (e); $R_{\protect\phi }=0.2$, $L_{%
\protect\phi }=0.4$ (f).}
\label{degenerate_CSs}
\end{figure}

Two isomers of solitons of the SV type, with right- and left-handed
chiralities, are defined, respectively, by vorticity sets $\left(
S_{+},S_{-}\right) =\left( 0,+1\right) $ and $\left( \bar{S}_{+},\bar{S}%
_{-}\right) =\left( -1,0\right) $ in the two components. Numerical solutions
for them are produced, severally, by the following inputs:%
\begin{equation}
\phi _{+}^{(0)}=A_{1}\exp \left( -\alpha _{1}r^{2}\right) ,\phi
_{-}^{(0)}=A_{2}r\exp \left( i\theta -\alpha _{2}r^{2}\right) ,
\label{ansatz}
\end{equation}%
\begin{equation}
\overline{\phi _{+}^{(0)}}=-A_{2}r\exp \left( -i\theta -\alpha
_{2}r^{2}\right) ,\overline{\phi _{-}^{(0)}}=A_{1}\exp \left( -\alpha
_{1}r^{2}\right) ,  \label{ansatz2}
\end{equation}%
which are written in polar coordinates $\left( r,\theta \right) $, with real
$A_{1,2}$ and $\alpha _{1,2}>0$. These inputs are natural, as Eqs. (\ref%
{phi+}) and (\ref{phi-}) are fully compatible with the substitution of SV
\textit{ans\"{a}tze} with opposite chiralities \cite{Sakaguchi_2014},
\begin{eqnarray}
\left\{ \phi _{+},\phi _{-}\right\} _{\mathrm{SV}} &=&e^{-i\mu t}\left\{
f(r),e^{i\theta }g\left( r\right) \right\} ,  \label{SVansatz1} \\
\left\{ \overline{\phi _{+}},\overline{\phi _{-}}\right\} _{\mathrm{SV}}
&=&e^{-i\mu t}\left\{ -e^{-i\theta }g\left( r\right) ,f\left( r\right)
\right\} ,  \label{SVansatz2}
\end{eqnarray}%
where radial form-factors $f(r),g(r)$ obey equations
\begin{gather}
\left[ \mu +\frac{1}{2}\left( \frac{d^{2}}{dr^{2}}+\frac{1}{r}\frac{d}{dr}%
\right) +\left( f^{2}+g^{2}\right) \right] f  \notag \\
-\left( \frac{1}{r}g+\frac{dg}{dr}\right) =0,  \label{f}
\end{gather}%
\begin{gather}
\left[ \mu +\frac{1}{2}\left( \frac{d^{2}}{dr^{2}}+\frac{1}{r}\frac{d}{dr}%
\right) +\left( f^{2}+g^{2}\right) \right] g  \notag \\
-\left( \frac{1}{2r^{2}}g-\frac{df}{dr}\right) =0.  \label{g}
\end{gather}%
More general soliton modes are introduced below.

\subsection{Exact Bessel modes of the linearized single-layer system}

Equations (\ref{f}) and (\ref{g}) give rise to solitons if $\mu $ belongs to
the respective \textit{semi-infinite bandgap}, $\mu <-1/2$ [see Eq. (\ref{mu}%
) below, with $\kappa =0$]. At $r\rightarrow \infty $, the linearization of
Eqs. (\ref{f}) and (\ref{g}) readily predicts that exponentially decaying
tails of solitons' form-factors $f(g)$ and $g(r)$ are built as combinations
of terms%
\begin{equation}
\left( f,g\right) _{\mathrm{tail}}\sim r^{-1/2}\exp \left( -\sqrt{-\left(
2\mu +1\right) }r\right) \left\{ \cos r,\sin r\right\} .  \label{tails}
\end{equation}%
Furthermore, the linearized equations make it possible to find exact
solutions for weakly localized Bessel-shaped modes (with a diverging total
norm) in the propagation band (at $\mu >-1/2$), which resemble exact
nondiffracting states in linear optics, cf. Refs. \cite{nondiffr0}-\cite%
{nondiffr2}). At $\mu >0$, the solutions of the SV type are%
\begin{equation}
f_{\mathrm{lin}}(r)=A^{(0)}J_{0}\left( r/\rho \right) ,~g_{\mathrm{lin}%
}(r)=A^{(0)}sJ_{1}\left( r/\rho \right)  \label{fg}
\end{equation}%
with arbitrary amplitude $A^{(0)}$, Bessel functions $J_{0,1}$, sign factor $%
s=\pm 1$ (this sign is independent from subscripts in $\phi _{\pm }$,$\psi
_{\pm }$), and
\begin{equation}
\rho =\rho _{s}\equiv \left( \sqrt{1+2\mu }+s\right) /\left( 2\mu \right) .
\label{rho1}
\end{equation}%
In the remaining part of the propagation band, $-1/2<\mu <0$, solution (\ref%
{fg}) is relevant only for $s=-1$, while expression (\ref{rho1}) is replaced
by
\begin{equation}
\rho =\rho _{\pm }\equiv \left( 1\pm \sqrt{1+2\mu }\right) /\left( -2\mu
\right) ,  \label{rho2}
\end{equation}%
with another independent sign $\pm $.

At the edge of the semi-infinite gap, $\mu =-1/2$, Eq. (\ref{rho2}) gives a
single solution with $\rho =1$,%
\begin{eqnarray}
f_{\mathrm{lin}}^{(1)}\left( r;\mu =-\frac{1}{2}\right)
&=&A^{(0)}J_{0}\left( r\right) ,~  \notag \\
g_{\mathrm{lin}}^{(1)}\left( r;\mu =-\frac{1}{2}\right)
&=&-A^{(0)}J_{1}\left( r\right) ,~  \label{1}
\end{eqnarray}%
instead of two, produced inside the gap at $\mu >-1/2$. However, the loss of
one solution is compensated by the existence of an additional one, precisely
at the edge of gap. Its components are%
\begin{eqnarray}
f_{\mathrm{lin}}^{(2)}\left( r;\mu =-\frac{1}{2}\right) &\equiv &\lim_{\mu
+1/2\rightarrow +0}\frac{f_{\mathrm{lin}}(r;\rho _{+})-f_{\mathrm{lin}%
}(r;\rho _{-})}{2\sqrt{1+2\mu }}  \notag \\
&=&-A^{(0)}rJ_{1}(r),  \notag \\
g_{\mathrm{lin}}^{(2)}\left( r;\mu =-\frac{1}{2}\right) &\equiv &\lim_{\mu
+1/2\rightarrow +0}\frac{g_{\mathrm{lin}}(r;\rho _{+})-g_{\mathrm{lin}%
}(r;\rho _{-})}{2\sqrt{1+2\mu }}  \notag \\
&=&A^{(0)}\left[ rJ_{2}(r)-J_{1}(r)\right] .  \label{2}
\end{eqnarray}

Exact linear SV states with the opposite chirality are generated from
expressions (\ref{fg})-(\ref{2}) as per Eq. (\ref{SVansatz2}). Next, MMs
states, as exact solutions of the linear SOC system, are constructed from
the SVs pursuant to Eqs. (\ref{MMansatz}) and (\ref{MMansatz2}). It is also
relevant to mention that, for the Rashba SOC\ replaced by the Dresselhaus
form \cite{Dressel}, the exact linear solutions for the SV and MM modes are
essentially the same as found here for the Rashba system.

In addition to solution (\ref{fg}), which represents the basic SV\ state in
the linear system, it is possible to construct exact solutions for excited
states, obtained by injection of extra integer vorticity $\Delta S=1,2,...$
in both components of the SV state. Such solutions are obtained replacing
Eqs. (\ref{fg}) and (\ref{1}), (\ref{2}), severally, by%
\begin{equation}
f_{\mathrm{lin}}(r)=A^{(0)}J_{\Delta S}\left( r/\rho \right) ,~g_{\mathrm{lin%
}}(r)=A^{(0)}sJ_{1+\Delta S}\left( r/\rho \right) ,  \label{fg-excited}
\end{equation}%
\begin{eqnarray}
f_{\mathrm{lin}}^{(1)}\left( r;\mu =-\frac{1}{2}\right) &=&A^{(0)}J_{\Delta
S}\left( r\right) ,~  \notag \\
g_{\mathrm{lin}}^{(1)}\left( r;\mu =-\frac{1}{2}\right)
&=&-A^{(0)}J_{1+\Delta S}(r),~  \label{1-excited}
\end{eqnarray}%
\begin{gather}
f_{\mathrm{lin}}^{(2)}\left( r;\mu =-\frac{1}{2}\right)  \notag \\
=A^{(0)}\left[ \Delta S\cdot J_{\Delta S}(r)-rJ_{1+\Delta S}(r)\right] ,
\notag \\
g_{\mathrm{lin}}^{(2)}\left( r;\mu =-\frac{1}{2}\right)  \notag \\
=A^{(0)}\left[ rJ_{2+\Delta S}(r)-\left( 1+\Delta S\right) J_{1+\Delta S}(r)%
\right] ,  \label{2-excited}
\end{gather}%
while Eqs. (\ref{rho1}) and (\ref{rho2}) keep the same form as above, as
well as the possibility to construct the SV with the opposite chirality and
MMs, according to Eqs. (\ref{SVansatz2}) and (\ref{MMansatz}), (\ref%
{MMansatz2}).

It is relevant to mention that similarly defined excited states of SV and MM
solitons are unstable in the nonlinear system, unlike the ground states with
$\Delta S=0$ \cite{Sakaguchi_2014}. On the other hand, the excited-state
solitons may be made stable in a model with self-trapping provided by the
local repulsive nonlinearity growing fast enough from the center to
periphery \cite{frontiers}.

\subsection{Soliton families}

The Manakov's nonlinearity makes it possible to construct a general family
of solitons, including SVs, MMs, and intermediate states. The creation of
such states is initiated by a combination of inputs (\ref{ansatz}) and (\ref%
{ansatz2}) with the opposite chiralities,%
\begin{equation}
\phi _{\pm }=M\phi _{\pm }^{(0)}+\sqrt{1-M^{2}}\cdot \overline{\phi _{\pm
}^{(0)}},  \label{M}
\end{equation}%
where the weight factor takes values $-1\leq M\leq 1$. Because integrals of
cross-products, generated by \textit{ans\"{a}tze} (\ref{ansatz}) and (\ref%
{ansatz2}) vanish,%
\begin{equation}
\int \int \left( \phi _{+}^{(0)}\right) ^{\ast }\phi _{-}^{(0)}dxdy=0,
\label{ortho}
\end{equation}%
the norm of input (\ref{M}), defined as per Eq. (\ref{N}), does not depend
on $M$, hence this input gives rise to a family of solitons with the same
norm.

The choice of $M=+1/\sqrt{2}$ or $-1/\sqrt{2}$ in Eq. (\ref{M}) corresponds
to MMs with opposite chiralities. Stationary MM solitons of the Manakov
system are produced by the following \emph{exact} ansatz, which is
compatible with the underlying system of equations (\ref{phi+}) and (\ref%
{phi-}) [cf. Eq. (\ref{SVansatz1}), that produces the exact ansatz for SV
modes]:%
\begin{gather}
\left\{ \phi _{+},\phi _{-}\right\} _{\mathrm{MM}}  \notag \\
=\frac{1}{\sqrt{2}}e^{-i\mu t}\left\{ f\left( r\right) -e^{-i\theta }g\left(
r\right) ,f\left( r\right) +e^{i\theta }g\left( r\right) \right\}   \notag \\
\equiv \frac{1}{\sqrt{2}}\left( \left\{ \phi _{+},\phi _{-}\right\} _{%
\mathrm{SV}}+\left\{ \overline{\phi _{+}},\overline{\phi _{-}}\right\} _{%
\mathrm{SV}}\right) .  \label{MMansatz}
\end{gather}%
Here $f$ and $g$ are the same real functions which are introduced above as
solutions to Eqs. (\ref{f}) and (\ref{g}). Another chiral MM\ isomer is
produced by a different ansatz,%
\begin{gather}
\left\{ \overline{\phi _{+}},\overline{\phi _{-}}\right\} _{\mathrm{MM}}=
\notag \\
\frac{1}{\sqrt{2}}e^{-i\mu t}\left\{ f\left( r\right) +e^{-i\theta }g\left(
r\right) ,-f\left( r\right) +e^{i\theta }g\left( r\right) \right\}   \notag
\\
\equiv \frac{1}{\sqrt{2}}\left( \left\{ \phi _{+},\phi _{-}\right\} _{%
\mathrm{SV}}-\left\{ \overline{\phi _{+}},\overline{\phi _{-}}\right\} _{%
\mathrm{SV}}\right) ,  \label{MMansatz2}
\end{gather}%
with radial form-factors $f$ and $g$ obeying the same equations (\ref{f})
and (\ref{g}). The exact linear relations between the MM and SVs in Eqs. (%
\ref{MMansatz}) and (\ref{MMansatz2}) are admitted by the symmetry of the
Manakov nonlinearity. These relations entail an equality for the total norms
(\ref{N}): $N_{\mathrm{MM}}(\mu )=N_{\mathrm{SV}}(\mu )$. To the best of our
knowledge, the exact MM \textit{ans\"{a}tze }represented by Eqs. (\ref%
{MMansatz})-(\ref{MMansatz2}) were not reported earlier.

The solitons are characterized by the norm ratio between the components,%
\begin{equation}
R_{\phi }=(N_{\phi _{+}}-N_{\phi _{-}})/(N_{\phi _{+}}+N_{\phi _{-}}),
\label{R}
\end{equation}%
and by the angular momentum per particle in each component and in the entire
system,%
\begin{eqnarray}
L_{\pm } &=&N_{\pm }^{-1}\int \int \phi _{\pm }^{\ast }\hat{\mathcal{L}_{z}}%
\phi _{\pm }dxdy,  \label{L} \\
L_{\phi } &=&N_{\phi }^{-1}\sum_{+,-}\int \int \phi _{\pm }^{\ast }\hat{%
\mathcal{L}_{z}}\phi _{\pm }dxdy,  \label{LL}
\end{eqnarray}%
where the angular-momentum operator is $\hat{\mathcal{L}_{z}}=-i(x\partial
/\partial y-y\partial /\partial x)\equiv -i\partial /\partial \theta $, and
the norms are taken as per Eq. (\ref{N}). For the family generated by ansatz
(\ref{M}), a typical relation between $L$ and $R_{\phi }$ is displayed in
Fig. \ref{degenerate_CSs}(a). The linear $L_{\phi }\left( R_{\phi }\right) $
dependence, observed in the figure, is explained by the fact that the input
in the form of Eqs. (\ref{ansatz})-(\ref{M}), leads, taking into regard the
orthogonality of $\phi _{+}^{(0)}$ and $\phi _{-}^{(0)}$ [see Eq. (\ref%
{ortho})], to $R_{\phi }(M)=\left( 2M^{2}-1\right) R_{\phi }^{\mathrm{(SV)}}$
and%
\begin{equation}
~L_{\phi }(R_{\phi })=\left( 1-R_{\phi }^{\mathrm{(SV)}}\right) \left(
2R_{\phi }^{\mathrm{(SV)}}\right) ^{-1}R_{\phi },  \label{linear}
\end{equation}%
where $R_{\phi }^{\mathrm{(SV)}}$ is given by Eq. (\ref{R}) for the SV
soliton corresponding to $M=1$ in Eq. (\ref{M})) [in Fig. \ref%
{degenerate_CSs}(a), a numerically computed value is $R_{\phi }^{\mathrm{(SV)%
}}=0.2$].

Both $R_{\phi }$ and $L_{\phi }$ vanish for solitons of the MM type, due to
the overall symmetry between the components in this state. Terminal points
of the curves in Fig. \ref{degenerate_CSs}(a) correspond to $|M|=1$ or $M=0$
in Eq. (\ref{M}), i.e., SV solitons, in which either $L_{+}=\pm 1$ and $%
L_{-}=0$, or vice versa. Numerical data yields $L_{\phi }=\pm 0.4$ in Eq. (%
\ref{linear}) at the terminal points. In other states, which correspond to $%
0<|M|<1$ in Eq. (\ref{M}), the numerical results give $-0.4<L_{\phi }<0.4$.
Examples of solitons belonging to the family are presented in Figs. \ref%
{degenerate_CSs}(b-f). In particular, SVs are shown in panels (b) and (f),
while (d) displays an MM. The stability of the entire family was established
in direct simulations. The solitons of all types exist in ranges $0<-\mu
<\infty $ and $0<N_{\phi }<N_{\text{\textrm{coll}}}\approx 5.85$, where $N_{%
\text{\textrm{coll}}}$ is the well-known threshold for the onset of the
critical collapse, predicted by the single Gross-Pitaevskii equation \cite%
{Fibich}. In fact, the stability of the 2D spin-orbit-coupled solitons is
predicated on the fact that their total norm takes values $N<N_{\text{%
\textrm{coll}}}$, thus securing them against the onset of the collapse \cite%
{Sakaguchi_2014}.

\subsection{The Josephson-coupled dual-core system}

The pair of Josephson-coupled layers, each carrying pseudo-spinor BEC, are
modeled by the system of equations for four components of wave functions, $%
\phi _{\pm }$ in one layer and $\psi _{\pm }$ in the other:%
\begin{gather}
i\partial _{t}\phi _{+}=-\left[ \frac{1}{2}\nabla ^{2}+\left( |\phi
_{+}|^{2}+\gamma |\phi _{-}|^{2}\right) \right] \phi _{+}  \notag \\
+\left( \partial _{x}-i\partial _{y}\right) \phi _{-}-\kappa \psi _{+},
\label{phi++}
\end{gather}%
\begin{gather}
i\partial _{t}\phi _{-}=-\left[ \frac{1}{2}\nabla ^{2}+\left( |\phi
_{-}|^{2}+\gamma |\phi _{+}|^{2}\right) \right] \phi _{-}  \notag \\
-\left( \partial _{x}+i\partial _{y}\right) \phi _{+}-\kappa \psi _{-},
\label{phi--}
\end{gather}%
\begin{gather}
i\partial _{t}\psi _{+}=-\left[ \frac{1}{2}\nabla ^{2}+\left( |\psi
_{+}|^{2}+\gamma |\psi _{-}|^{2}\right) \right] \psi _{+}  \notag \\
+\left( \partial _{x}-i\partial _{y}\right) \psi _{-}-\kappa \phi _{+},
\label{psi++}
\end{gather}%
\begin{gather}
i\partial _{t}\psi _{-}=-\left[ \frac{1}{2}\nabla ^{2}+\left( |\psi
_{-}|^{2}+\gamma |\psi _{+}|^{2}\right) \right] \psi _{-}  \notag \\
-\left( \partial _{x}+i\partial _{y}\right) \psi _{+}-\kappa \phi _{-},
\label{psi--}
\end{gather}%
cf. Eqs. (\ref{phi+}) and (\ref{phi-}), where $\kappa $ is the
Josephson-coupling strength, which is defined to be positive. Note that the
Josephson interaction does not linearly couple components $\phi _{\pm }$ to $%
\psi _{\mp }$, as ones with opposite subscripts represent two distinct
atomic states, and the tunneling between the parallel layers does not
include a mechanism which would lead to mutual transformation of different
atomic states.

The system of Eqs. (\ref{phi++})-(\ref{psi--}) conserves the total norm, $%
N_{\phi }+N_{\psi }$, where $N_{\psi }$ is a counterpart of the $\phi $ norm
for the $\psi $-layer, see Eq. (\ref{N}). Other dynamical invariants of the
system are the total energy,
\begin{gather}
E=\iint \left\{ \sum_{\chi =\phi ,\psi }\left[ \frac{1}{2}\left( \left\vert
\nabla \chi _{+}\right\vert ^{2}+\left\vert \nabla \chi _{-}\right\vert
^{2}-\left\vert \chi _{+}\right\vert ^{4}-\left\vert \chi _{-}\right\vert
^{4}\right) \right. \right.  \notag \\
\left. -\gamma \left\vert \chi _{+}\chi _{-}\right\vert ^{2}+\left( \left(
\chi _{+}^{\ast }\partial _{x}\chi _{-}-i\chi _{+}^{\ast }\partial _{y}\chi
_{-}\right) +\mathrm{c.c.}\right) \right]  \notag \\
\left. -\kappa \left( \left( \phi _{+}^{\ast }\psi _{+}+\phi _{-}^{\ast
}\psi _{-}\right) +\mathrm{c.c.}\right) \right\} dxdy,  \label{E}
\end{gather}%
and the total angular momentum, $N_{\phi }L_{\phi }+N_{\psi }L_{\psi }$, cf.
Eq. (\ref{LL}).

\section{Analytical and numerical results for Josephson oscillations}

\subsection{Exact solutions for Josephson oscillations in the linear system}

Linearization of Eqs. (\ref{phi++})-(\ref{psi--}) for excitations $\left(
\phi _{\pm },\psi _{\pm }\right) \sim \exp \left( iq_{x}x+iq_{y}y-i\mu
t\right) $ yields four branches of the dispersion relation:
\begin{equation}
\mu =\frac{1}{2}\left( q_{x}^{2}+q_{y}^{2}\right) +\sigma _{1}\sqrt{%
q_{x}^{2}+q_{y}^{2}}-\sigma _{2}\kappa ,  \label{mu}
\end{equation}%
with two mutually independent sign parameters $\sigma _{1,2}=\pm 1$ (which
are unrelated to subscripts of $\phi _{\pm }$ and $\psi _{\pm }$). Solitons
may exist in the semi-infinite gap, readily determined as an interval of
values of $\mu $ which is not covered by Eq. (\ref{mu}) with $-\infty
<q_{x,y}<+\infty $ and all choices of $\sigma _{1,2}$:
\begin{equation}
\mu _{\mathrm{sol}}<-\left( 1/2+\kappa \right)  \label{SIG}
\end{equation}%
(remind we fix $\kappa >0$, by definition). In particular, this gap is
obtained from the above-mentioned one, $\mu _{\mathrm{sol}}<-1/2$, for the
single-layer system, if one considers symmetric solutions of Eqs. (\ref%
{phi++})-(\ref{psi--}), with $\psi _{\pm }=\phi _{\pm }$. On the other hand,
for antisymmetric solutions, with $\psi _{\pm }=-\phi _{\pm }$, the
single-layer gap is replaced by an expanded one, $\mu _{\mathrm{sol}%
}<-\left( 1/2-\kappa \right) $. In the additional interval which does not
belong to gap (\ref{SIG}), \textit{viz}., $-\left( 1/2+\kappa \right) <\mu _{%
\mathrm{sol}}<-\left( 1/2-\kappa \right) $, antisymmetric states may exist
as \textit{embedded solitons} \cite{embedded}, but they are unstable.

Further, the linearized system gives rise to \emph{exact }solutions for
Josephson-oscillation states,%
\begin{gather}
\left( \phi _{\pm }\left( x,y,t\right) \right) _{\mathrm{two-layer}}=\cos
\left( \kappa t\right) \cdot \left( \phi _{\pm }\left( x,y,t\right) \right)
_{\mathrm{single-layer}},  \notag \\
\left( \psi _{\pm }\left( x,y,t\right) \right) _{\mathrm{two-layer}}=i\sin
\left( \kappa t\right) \cdot \left( \phi _{\pm }\left( x,y,t\right) \right)
_{\mathrm{single-layer}},  \label{exact}
\end{gather}%
where $\left( \phi _{\pm }\left( x,y,t\right) \right) _{\mathrm{single-layer}%
}$ are the component of any solution of the linearized version of Eqs. (\ref%
{phi+}) and (\ref{phi-}). In fact, explicit single-layer solutions which may
be substituted in Eq. (\ref{exact}) are those given above by Eqs. (\ref{fg}%
)-(\ref{2-excited}). Note that the Josephson oscillations do not change the
chemical potential in Eq. (\ref{exact}), which is borrowed from the
respective single-layered solutions, in the form of $\exp \left( -i\mu
t\right) $.

\subsection{Josephson oscillations in 2D solitons}

\subsubsection{Chirality and identity oscillations}

In the framework of the full nonlinear system, we first address the
evolution initiated by an input composed of a stable single-layer SV
soliton, with vorticities $\left( S_{+},S_{-}\right) =\left( 0,1\right) $,
in one core, and its chiral (left-handed) isomer with $\left(
S_{+},S_{-}\right) =\left( -1,0\right) $ in the other, which are created, at
$t=0$, as per \textit{ans\"{a}tze} presented in Eqs. (\ref{SVansatz1}) and (%
\ref{SVansatz2}). The respective stationary single-layer solutions were
produced by means of the accelerated imaginary-time-evolution method \cite%
{yang2010nonlinear}, and their stability was then tested by simulations of
Eqs. (\ref{phi+}) and (\ref{phi-}) in real time.

\begin{figure}[tbp]
\centering{\label{fig2a} \includegraphics[scale=0.19]{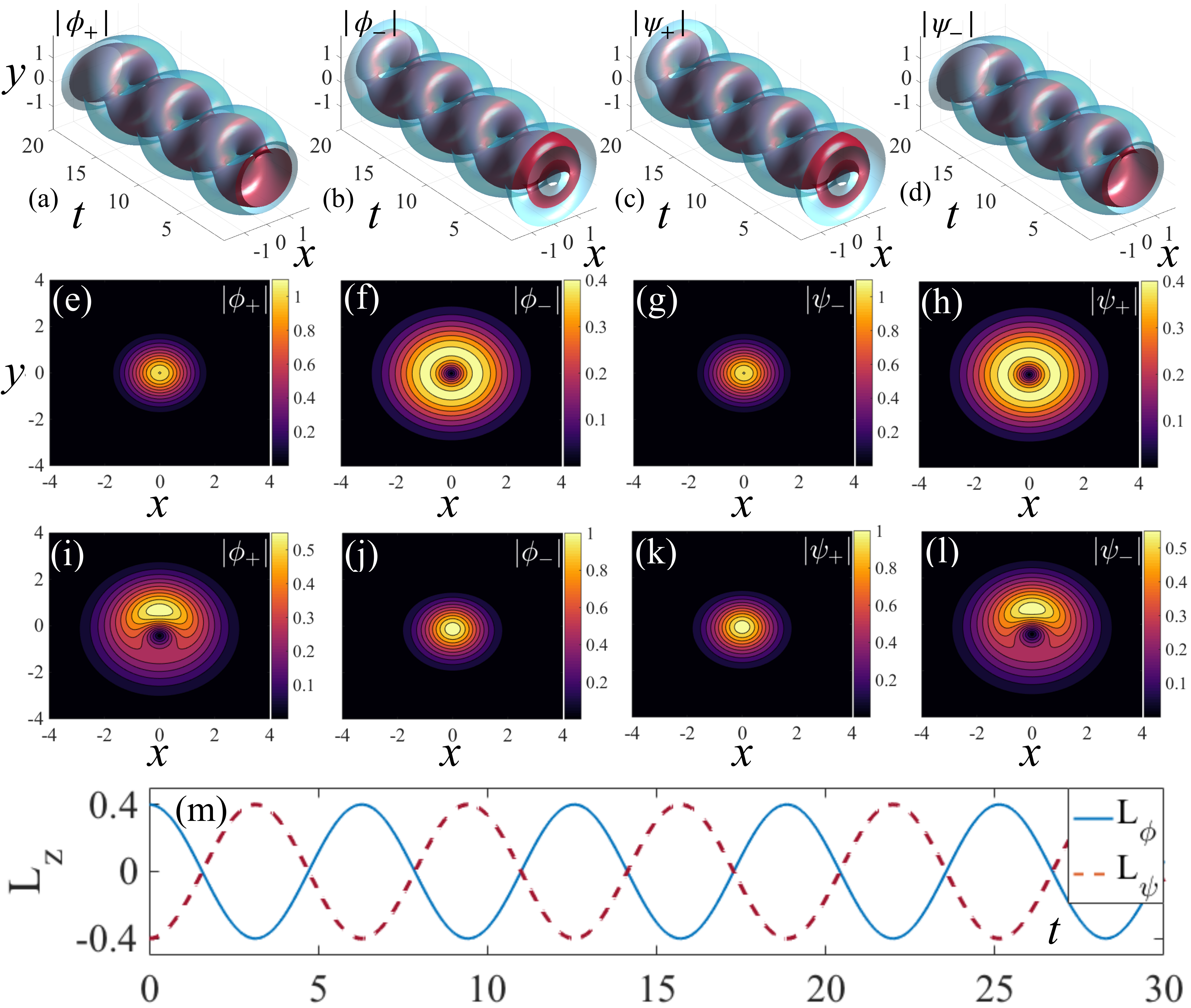}}
\caption{(Color online) Numerically simulated Josephson oscillations of
chirality (with random-noise perturbation at the $1\%$ amplitude level,
added for testing stability of the dynamical regime), initiated by the input
composed of a stable single-core SV in one layer, and its left-handed
counterpart in the other, see Eqs. (\protect\ref{SVansatz1}) and (\protect
\ref{SVansatz2}). Panels (a,b) and (c,d) display the evolution of the
pseudo-spinor components in the first and second cores, respectively. The
norm of the input in each layer is $N=4$ (hence the total norm in the
two-layer system is $8$), and the Josephson coupling constant is $\protect%
\kappa =0.5$. Panel (e)-(h) and (i)-(l) exhibit 2D density distributions of
the components at, respectively, $t=0$ and $t=10$. (m) The respective
oscillations of angular momenta per particle in both layers , $L_{\protect%
\phi }$ and $L_{\protect\psi }$, see Eq. (\protect\ref{LL}).}
\label{JOSM_SV_ISV_DR}
\end{figure}

Simulations of Eqs. (\ref{phi++})-(\ref{psi--}) with such an input produce a
robust regime of periodic transmutations displayed in Fig. \ref%
{JOSM_SV_ISV_DR}(a-d). Each quarter period, SVs in both cores switch into
their chiral counterparts, returning to the initial state, but with the
opposite sign, each half period. Thus, chiralities of the two SVs
periodically flip, remaining opposite in the two cores, while the
corresponding density patterns oscillate. Distributions of densities of the
four involved components, displayed at $t=0$ in panels (e-f), and at $t=10$
in (i-l) (the latter time is close to $0.8$ periods of the Josephson
oscillations) help to understand the evolution of the density patterns in
the course of the oscillations. Further, Fig. \ref{JOSM_SV_ISV_DR}(m)
demonstrates Josephson oscillations of the angular momentum in the same
dynamical state, with the momentum per particle in the cores oscillating
between $L_{\phi ,\psi }=\pm 0.4$ and $\mp 0.4$.

Note that the chirality-switching Josephson-oscillation mode is possible
only in the nonlinear system, as the exact linear Josephson-oscillation
solution (\ref{exact}) admits solely oscillations that do not mix different
chiralities. Indeed, the chirality flipping implies generation of new
angular harmonics, which is not possible without nonlinearity.

\begin{figure}[tbp]
\centering{\label{fig5a} \includegraphics[scale=0.20]{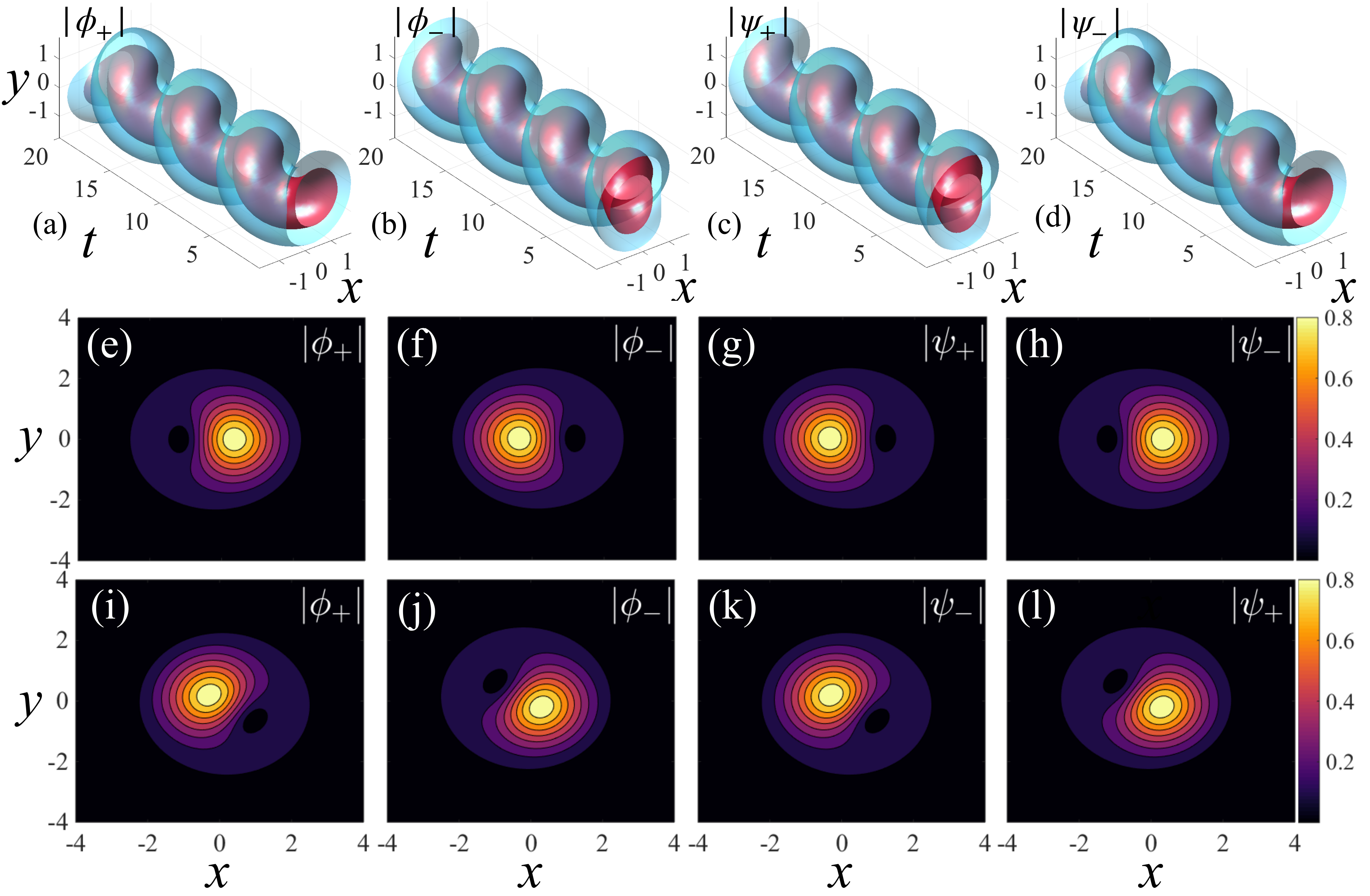}}
\caption{(Color online) The same as in Fig. \protect\ref{JOSM_SV_ISV_DR} but
for the input in the form of MMs with opposite chiralities in the two cores,
and equal norms in them, $N=4$. The angular momentum per particle remains
equal to zero in each core, due to the symmetry between the components of
the pseudo-spinor wave function in the MM states. In addition to the
Josephson oscillations of the MM chirality, the figure also demonstrates the
spatiotemporal helicity in the oscillating soliton, see the text. Panel
(e)-(h) and (i)-(l) exhibit 2D density distributions of the components at,
respectively, $t=0$ and $t=10$. }
\label{JOSM_MM_IMM_DR}
\end{figure}

Next, we consider Josephson oscillations between right- and left-handed
chiral isomers of the MM type, which are created, at $t=0$, in the two
layers, as per \textit{ans\"{a}tze} (\ref{MMansatz}) and (\ref{MMansatz2}).
A typical numerical solution, displayed in Fig. \ref{JOSM_MM_IMM_DR},
features stable periodic switching between the MM\ isomers in each layer. It
also demonstrates \textit{spatiotemporal helicity}, i.e., clockwise rotation
of the density distributions (unlike nonrotating SV\ patterns in Fig. \ref%
{JOSM_SV_ISV_DR}). The input with the opposite relative sign between the MM
isomers in two layers gives rise to a similar stable dynamical state, but
with rotation of the density profiles in the opposite, counter-clockwise,
direction. Again, it is relevant to stress that flipping of the MM chirality
in the course of the Josephson oscillations is only possible in the
nonlinear system.

Recall that the single-core SOC system with the Manakov nonlinearity
maintains the family of 2D solitons which are intermediate between the SV
and MM states. Such states correspond to
\begin{equation}
|M|\neq 0,\frac{1}{\sqrt{2}},1  \label{interm}
\end{equation}%
in Eq. (\ref{M}), $\left\vert R_{\phi }\right\vert \neq 0.2$ or $0$ in Fig. %
\ref{degenerate_CSs}, and accordingly, eigenvalues $L_{\phi }$ different
from $\pm 0.4$ and $0$, see Eq. (\ref{LL}). Systematic simulations of
Josephson oscillations in pairs of chiral isomers of the intermediate states
also reveal robust periodic chirality oscillations, in a combination with
the rotation of the density profiles (the spatiotemporal helicity), similar
to what is displayed above in Figs. \ref{JOSM_SV_ISV_DR} and \ref%
{JOSM_MM_IMM_DR} for the states of the SV and MM types. The results for the
intermediate states are not displayed in a separate figure, as the overall
pictures seem quite close to ones in Figs. \ref{JOSM_SV_ISV_DR} and \ref%
{JOSM_MM_IMM_DR}). Furthermore, at all values of inter-core coupling $\kappa
$ the frequency of the oscillations and rotation is exactly equal to $\kappa
$, cf. the exact solution of the linear system given by Eqs. (\ref{phi++})-(%
\ref{psi--}). Although, as said above, the chirality oscillations represent
a nonlinear effect, the Josephson frequency is not affected by the
nonlinearity, due to its \textquotedblleft isotopic" invariance with respect
to the rotation of the pseudo-spinor wave function in the plane of its two
components.

\begin{figure}[tbp]
\centering{\label{fig4a} \includegraphics[scale=0.185]{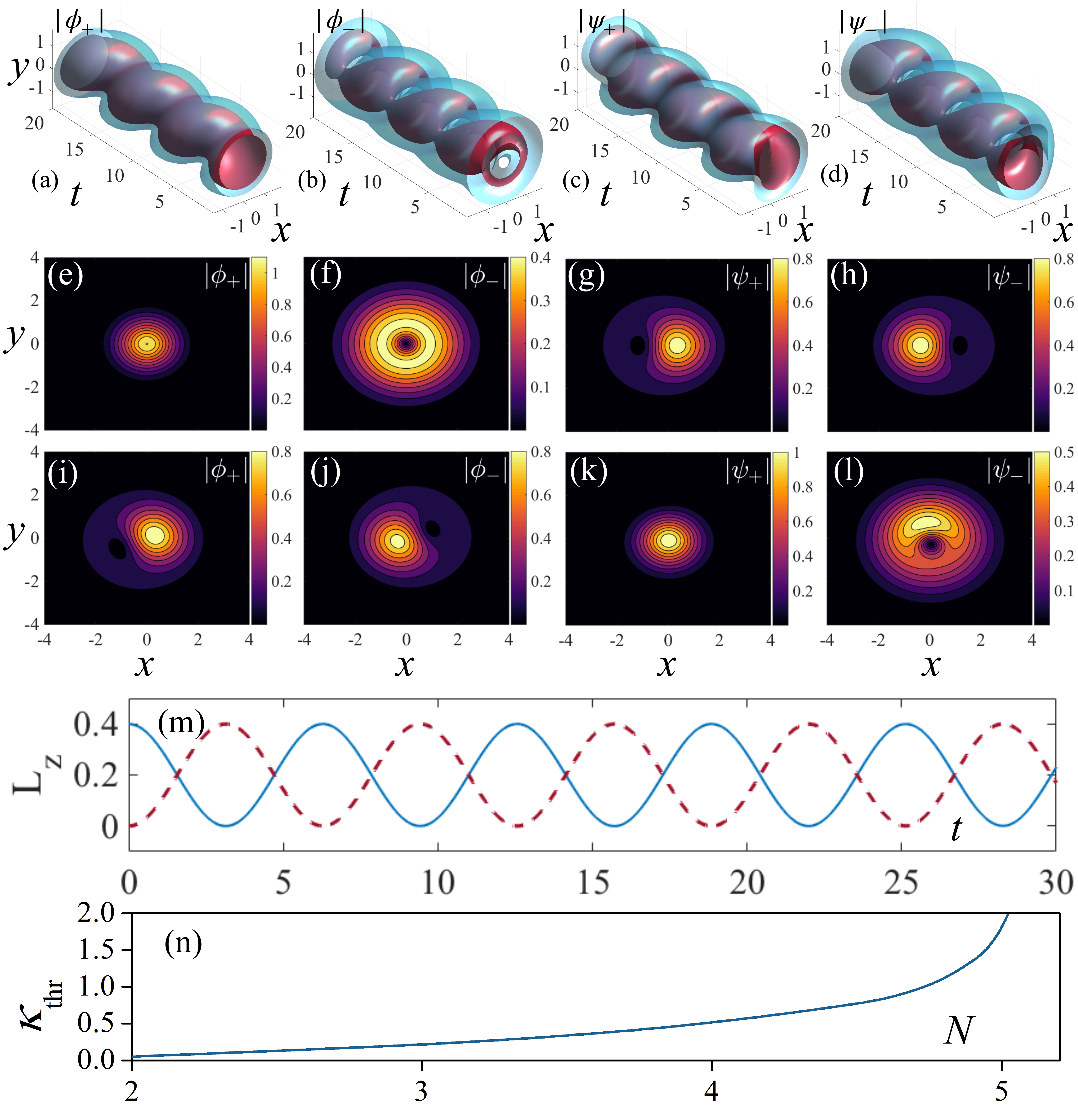}}
\caption{(Color online) Panels (a-m): the same as in Fig. \protect\ref%
{JOSM_SV_ISV_DR}, but for \textit{identity oscillations}, initiated by the
input in the form of SV and MM solitons with equal norms, $N=4$, in the
coupled cores. (n) The minimum (threshold) value of the inter-core coupling $%
\protect\kappa $, which is necessary for the stability of the identity
oscillations, vs. the norm of the 2D soliton in each core, $N$.}
\label{SV-MM}
\end{figure}

The next step is to consider Josephson oscillations between inputs
corresponding to SV in one core and MM, with the same norm, in the other. In
this case, the \textit{identity} of the 2D solitons periodically switches
between the SV and MM types in each core [see Figs. \ref{SV-MM}(a-d) and
(m)], therefore this stable dynamical state may be naturally categorized as
\textit{identity oscillations}, coupled to the clockwise helicity rotation,
cf. Figs. \ref{JOSM_SV_ISV_DR} and \ref{JOSM_MM_IMM_DR}. This regime, which
is again possible solely in the nonlinear system, cf. exact solution (\ref%
{exact}) obtained in the linear limit (which actually corresponds to the
system strongly dominated by the inter-core coupling), takes place if the
strength of the Josephson interaction exceeds a certain threshold value, $%
\kappa >\kappa _{\mathrm{thr}}$. The dependence of the threshold on the
total norm in one layer, $N$, is shown in Fig. \ref{SV-MM}(n); recall that
the establishment of the robust chirality oscillations in the dual-core SV
and MM solitons, considered above, does not require $\kappa $ to be larger
than any finite threshold value.

To explain the latter feature, we note that, as it follows from Eqs. (\ref%
{phi++})-(\ref{psi--}) and (\ref{E}), the interplay of the kinetic energy,
nonlinearity, SOC, and Josephson interaction between the cores predicts
scaling $\kappa _{\mathrm{thr}}\sim N$ at small $N$. On the other hand, the
above-mentioned onset of the collapse in the single-core system at $N=N_{%
\text{\textrm{coll}}}$ implies divergence of the system's sensitivity to
perturbations at $N\rightarrow N_{\text{\textrm{coll}}}$, hence $\kappa _{%
\mathrm{thr}}$ also diverges in this limit. These qualitative arguments are
corroborated by the numerically found dependence $\kappa _{\mathrm{thr}}(N)$
in Fig. \ref{SV-MM}(n).

At $\kappa <\kappa _{\mathrm{thr}}$, the inter-core coupling is too weak to
prevent the occurrence of spontaneous symmetry breaking between the cores.
Accordingly, additional numerical results demonstrate, for values of $N$
which are relatively small in comparison with $N_{\text{\textrm{coll}}}$
(namely, in an interval of $N<N_{\text{\textrm{crit}}}<N_{\text{\textrm{coll}%
}}$) the existence of asymmetric composite solitons, with different
amplitudes in the two cores. In the adjacent interval of $N_{\text{\textrm{%
crit}}}<N<N_{\text{\textrm{coll}}}$, the collapse takes place at $\kappa
<\kappa _{\mathrm{thr}}$. On the other hand, if the Josephson coupling is
strong enough, with $\kappa >\kappa _{\mathrm{thr}}$, it forestalls the
onset of the collapse at all values of $N$ smaller than $N_{\text{\textrm{%
coll}}}$, by keeping the total norm equally distributed between the two
cores (for instance, the identity-oscillation dynamical regime, displayed in
Fig. \ref{SV-MM}, is stable at $N=4<$ $N_{\text{\textrm{coll}}}\approx $ $%
5.85$).


\subsubsection{Structural stability of the oscillation regimes}

\begin{figure}[tbp]
\centering{\label{fig3a} \includegraphics[scale=0.3]{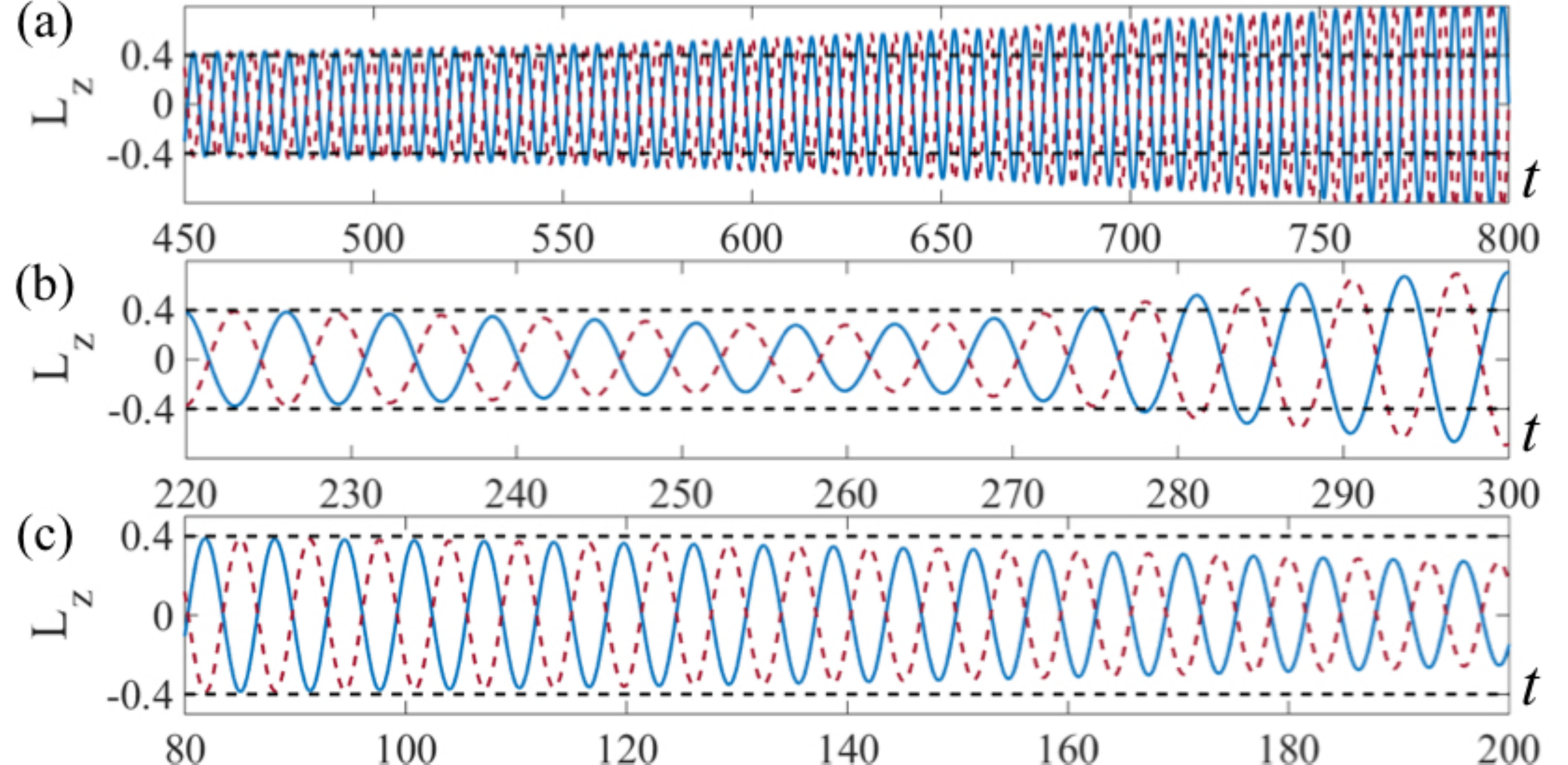}}
\caption{(Color online) Simulations similar to those displayed for the SV
soliton in Fig. \protect\ref{JOSM_SV_ISV_DR}, but with $\protect\gamma =1.05$
in (a) and $0.95$ in (b). The chirality oscillations remain robust in the
course of $\simeq 40$ and $20$ oscillation periods, respectively. (c) The
chirality oscillations of the SV solitons, with different initial norms in
the two cores, $N_{\protect\psi }=0.95N_{\protect\phi }=3.8$. The
oscillations stay undisturbed in the course of $\simeq 10$ periods.}
\label{JOSM_SV_MM_DR}
\end{figure}

The fact that the Manakov's form of the nonlinearity is approximate in
experimental settings makes it necessary to test effects of deviation from
it [$\gamma \neq 1$ in Eqs. (\ref{phi+}) and (\ref{phi-})], i.e., \textit{%
structural stability} of the chirality- and identity-oscillation dynamical
regimes. The result, illustrated by Figs. \ref{JOSM_SV_MM_DR}(a,b) for $%
\gamma =1\pm 0.05$, is that the chirality oscillations of the SV\ soliton
persist in the course of dozens of periods (which is sufficient for the
experimental observation), and then gradually detune from the regular regime.

Additional simulations (not shown here in detail) suggest similar
conclusions concerning the effect of the deviation from the Manakov's case
on chirality and identity oscillations of MM solitons. As concerns the
intermediate states, defined above as per Eq. (\ref{interm}), the deviation
from the Manakov's case for them is more disruptive, as, even in the
single-core system, taking $\gamma \neq 1$ causes evolution of such inputs
towards SVs or MMs, for $\gamma <1$ or $\gamma >1$, respectively \cite%
{Sakaguchi_2014}.

Stability of the Josephson oscillations against variations of the initial
conditions was tested too, by taking the input in the form of a single-core
SV soliton in one layer, and its chiral isomer in the other, but with
different norms. As shown in Fig. \ref{JOSM_SV_MM_DR}(c), dephasing of the
resulting chirality oscillations sets in at times exceeding $\sim 10$
periods.

\section{Conclusion}

The objective of this work is to introduce and theoretically elaborate a
setting for the study of Josephson oscillations of the chirality and species
type (\textit{identity}) of 2D solitons in the double-layer trap, loaded
with a spin-orbit-coupled condensate, including self- and cross-attractive
nonlinearity (SPM and XPM) in its components. It is known that, in the
Manakov's case of SPM = XPM, the single-layer system supports a broad family
of composite 2D solitons, which includes dynamically and structurally stable
ones of the SV (semi-vortex) and MM (mixed-mode) types, each existing in two
chiral isomers, left- and right-handed ones. In addition to that, the
Manakov's nonlinearity admits the existence of the intermediate family of
solitons, between the SV and MM states, which are dynamically stable but
structurally unstable against deviation from the condition SPM = XPM. All
these states exist with the norm in each component taking values smaller
than the well-known threshold value necessary for the onset of the critical
collapse, $N<N_{\mathrm{coll}}$,

We consider Josephson oscillations initiated by inputs in the form of SV,
MM, or intermediate-type solitons with equal norms and opposite chiralities,
originally created in the coupled cores (layers), or SV in one core and MM
in the other. In the former case, the system features robust chirality
oscillations at all values of the inter-core coupling, $\kappa $. In the
latter case, persistent periodic switching between the SV and MM species (%
\textit{identity oscillations}) occur at $\kappa >\kappa _{\mathrm{thr}}$,
where $\kappa _{\mathrm{thr}}$ increases with the growth of the condensate's
norm, while at $\kappa <\kappa _{\mathrm{thr}}$ the system develops states
with unequal amplitudes in the cores at relatively small values of $N$, or
collapses at larger $N$. In all cases, the oscillation frequency is equal to
$\kappa $. Deviation of the nonlinearity from the Manakov's form, as well as
a difference in the norms of the inputs in the two cores, leads to slowly
developing dephasing of the Josephson oscillations. These regimes are
nonlinear ones, while the exact solutions reported here for Josephson
oscillations in the linear system admit solely self-oscillations of SVs and
MMs modes with the Bessel spatial shape, without switching between different
chiralities or mode types. In fact, the exact Bessel-shaped linear modes are
new stationary solutions for the single-layer system too.

With typical values of the relevant physical parameters (see, e.g., Ref.
\cite{Romania}), such as the scattering length of the attractive
inter-atomic interactions $\sim -0.1$ nm, transverse trapping frequency $%
\sim 300$ Hz, and the gap of width $\sim 5$ $\mathrm{\mu }$m separating two
layers, the expected number of atoms in the stable solitons, which are
capable to feature the Josephson oscillations, is in the range of $N_{%
\mathrm{atom}}\sim 10^{3}-10^{4}$, and the oscillation frequency $\sim 20$
Hz (while the experiment time may extend up to $1$ s, making the observation
of many cycles of the oscillations quite feasible). At essentially larger
values of $N_{\mathrm{atom}}$, the scaled norm will exceed value $N_{\mathrm{%
coll}}$, which will trigger the onset of the collapse. On the other hand, it
was recently predicted \cite{Petrov1,Petrov2}, and demonstrated
experimentally in various setups \cite{exp1}-\cite{exp5}, that 3D and
quasi-2D self-trapped matter-wave states may be stabilized, in the form of
soliton-like \textit{quantum droplets}, by beyond-mean-field
(Lee-Huang-Yang) effects of quantum fluctuations around the mean-field
states. In the present context, this possibility suggests to consider the
dual-core SOC system in the case of $N>N_{\mathrm{coll}}$, that may be
stabilized by the Lee-Huang-Yang effect. In this connection, it is relevant
to mention that the stabilization of 2D spin-orbit-coupled solitons by this
effect in the single-layer system was analyzed in Ref. \cite{we}.

As another extension of the present work, it may be interesting to consider
Josephson oscillations in 2D spatiotemporal optical solitons carried by
dual-core planar waveguides and stabilized by a SOC-emulating mechanism,
\textit{viz}., temporal dispersion of the inter-core coupling coefficient
\cite{OptLett,Elad}. Another potentially interesting point is the study of
intrinsic vibrations of the composite solitons, initiated by a separation of
their components in the two layers.

\section*{Acknowledgments}

This work was supported, in part, by grants No.11874112 and No.11575063 from
NNSFC (China), and by grant No. 1286/17 from the Israel Science Foundation.
Z.C. acknowledges an excellence scholarship provided by the Tel Aviv
University.


\begin{thebibliography}{99}
\bibitem{book} B. A. Malomed (editor), \textit{Spontaneous Symmetry
Breaking, Self-Trapping, and Josephson Oscillations} (Springer, Berlin,
2013).

\bibitem{josephson1962} B. D. Josephson. \newblock Possible new effects in
superconductive tunnelling. \newblock {\em Physics letters}, \textbf{1},
251-253 (1962).

\bibitem{Anderson} P. W. Anderson and J. M. Rowell, Probable observation of
the Josephson superconducting tunneling effect, Phys. Rev. Lett. \textbf{10}%
, 230-232 (1963).

\bibitem{Barone} A. Barone and C. Paterno, \textit{Physics and Applications
of the Josephson Effect} (Wiley: New York, 1982).

\bibitem{Likharev} K. K. Likharev, \textit{Introduction to the Dynamics of
Josephson Junctions} (Nauka: Moscow, 1985) (in Russian).

\bibitem{Buzdin} A. I. Buzdin, Proximity effects in
superconductor-ferromagnet heterostructures, Rev. Mod. Phys. \textbf{77},
935-976 (2005).

\bibitem{Fulton} T. A. Fulton and R. C. Dynes, Single vortex propagation in
Josephson tunnel junctions, Solid State Commun. \textbf{12}, 57-61 (1973).

\bibitem{Cirillo} A. V. Ustinov, M. Cirillo, and B. A. Malomed, Fluxon
dynamics in one-dimensional Josephson-junction arrays, Phys. Rev. B \textbf{%
47}, 8357-8360 (1993).

\bibitem{Lomdahl} P. S. Lomdahl, Solitons in Josephson junctions -- an
overview, J. Stat. Phys. \textbf{39}, 551-561 (1985).

\bibitem{Ustinov} A. V. Ustinov, Solitons in Josephson junctions, Physica D
\textbf{123}, 315-329 (1998).

\bibitem{Walraff} A. Wallraff, A. Lukashenko, J. Lisenfeld, A. Kemp, M. V.
Fistul, Y. Koval, and A. V. Ustinov, Quantum dynamics of a single vortex,
Nature \textbf{425}, 155-158 (2003).

\bibitem{super} B. A. Malomed, V. A. Oboznov, and V. A. Ustinov,
Supersolitons in periodically inhomogeneous long Josephson junctions, Zh.
Eksp. Teor. Fiz. \textbf{97}, 924-937 (1990) [Sov. Phys. JETP \textbf{70},
518-525 (1990)].

\bibitem{davis2002superfluid} J. C.~Davis and R. E.~Packard, \newblock %
Superfluid $^{3}$He \ Josephson weak links.
\newblock {\em Reviews of Modern
Physics}, \textbf{74}, 741 (2002).

\bibitem{Milburn} G. J. Milburn, J. Corney, E. M. Wright, and D. F. Walls,
Quantum dynamics of an atomic Bose-Einstein condensate in a double-well
potential, Phys. Rev. A \textbf{55}, 4318-4324 (1997).

\bibitem{smerzi1997quantum} A. Smerzi, S. Fantoni, S. Giovanazzi, and S.
R.~Shenoy. \newblock Quantum coherent atomic tunneling between two trapped
Bose-Einstein condensates. \newblock {\em Physical Review Letters}, 79, 4950
(1997).

\bibitem{Raghavan} S. Raghavan, A. Smerzi, S. Fantoni, and S. R. Shenoy,
Coherent oscillations between two weakly coupled Bose-Einstein condensates:
Josephson effects, pi-oscillations, and macroscopic quantum self-trapping,
Phys. Rev. A \textbf{59}, 620-633 (1999).

\bibitem{albiez2005direct} M. Albiez, R. Gati, J. F{\"{o}}lling, S.
Hunsmann, M. Cristiani, and M. K. Oberthaler. \newblock Direct observation
of tunneling and nonlinear self-trapping in a single bosonic Josephson
junction. \newblock {\em Physical review letters}, 95, 010402 (2005)

\bibitem{Ketterle} Y. Shin, G.-B. Jo, M. Saba, T. A. Pasquini, W. Ketterle,
and D. E. Pritchard, Optical weak link between two spatially separated
Bose-Einstein condensates, Phys. Rev. Lett. \textbf{95}, 170402 (2005).

\bibitem{Steinhauer} S. Levy, E. Lahoud, I. Shomroni, and J. Steinhauer, The
a.c. and d.c. Josephson effects in a Bose--Einstein condensate, Nature
\textbf{449}, 579-583 (2007).

\bibitem{Salerno} G. Mazzarella, B. Malomed, L. Salasnich, M. Salerno, and
F. Toigo, Rabi-Josephson oscillations and self-trapped dynamics in atomic
junctions with two bosonic species, J. Phys. B: At. Mol. Opt. Phys. \textbf{%
44}, 035301 (2011).

\bibitem{Veronica} J. Polo, A. Benseny, Th. Busch, V. Ahufinger, and J.
Mompart, Transport of ultracold atoms between concentric traps via spatial
adiabatic passage, New J. Phys. \textbf{18}, 015010 (2016).


\bibitem{asymm1} C. Par\'{e} and M. F\l orja\'{n}czyk, Approximate model of
soliton dynamics in all-optical couplers, Phys. Rev. A \textbf{41},
6287-6295 (1990)

\bibitem{Maim} A. I. Maimistov, Propagation of a light pulse in nonlinear
tunnel-coupled optical waveguides, Kvant. Elektron. \textbf{18}, 758-761
[Sov. J. Quantum Electron. \textbf{21}, 687-690 (1991)].

\bibitem{Ivan} I. M. Uzunov, R. Muschall, M. Goelles, Yu. S. Kivshar, B. A.
Malomed, and F. Lederer, Pulse switching in nonlinear fiber directional
couplers, Phys. Rev. E \textbf{51}, 2527-2537 (1995).

\bibitem{Ignac} I. Bugar, I. V. Fedotov, A. B. Fedotov, M. Koys, R.
Buczynski, D. Pysz, J. Chlpik, F. Uherek, and A. M. Zheltikov,
Polarization-controlled dispersive wave redirection in dual-core photonic
crystal fiber, Laser Physics \textbf{18}, 1420-1428 (2008).

\bibitem{exciton} M. Abbarchi, A. Amo, V. G. Sala, D. D. Solnyshkov, H.
Flayac, L. Ferrier, I. Sagnes, E. Galopin, A. Lemaitre, G. Malpuech, and J.
Bloch, Macroscopic quantum self-trapping and Josephson oscillations of
exciton polaritons, Nature Phys. \textbf{5}, 275-279 (2013).

\bibitem{Levitov} J. E. Mooij, T. P. Orlando, L. Levitov, L. Tian, C. H. van
der Wal, and S. Lloyd,\ Josephson persistent-current qubit, Science \textbf{%
285}, 1036-1039 (1999).

\bibitem{Shnirman} Y. Makhlin, G. Sch\"{o}n, and A. Shnirman, Quantum-state
engineering with Josephson-junction devices, Rev. Mod. Phys. \textbf{73},
357-400 (2001).

\bibitem{lesanovsky2007spontaneous} I. Lesanovsky and W. von Klitzing. %
\newblock Spontaneous emergence of angular momentum Josephson oscillations
in coupled annular Bose-Einstein condensates.
\newblock {\em Physical review
letters}, 98, 050401 (2007).

\bibitem{Brand0} J. Brand, T. J. Haigh, and U. Z\"{u}licke, Sign of coupling
in barrier-separated Bose-Einstein condensates and stability of double-ring
systems, Phys. Rev. A 81, 025602 (2010).

\bibitem{Amico} D. Aghamalyan, L. Amico, and L. C. Kwek, Effective dynamics
of cold atoms flowing in two ring-shaped optical potentials with tunable
tunneling. Phys. Rev. A \textbf{88}, 063627 (2013).

\bibitem{Brand} S. W. Su, S. C. Gou, A. Bradley, O. Fialko, and J. Brand,
Kibble-Zurek scaling and its breakdown for spontaneous generation of
Josephson vortices in Bose-Einstein condensates. Phys. Rev. Lett. \textbf{110%
}, 215302 (2013).

\bibitem{Veronica2} J. Polo, J. Mompart, and V. Ahufinger, Geometrically
induced complex tunnelings for ultracold atoms carrying orbital angular
momentum, Phys. Rev. A \textbf{93}, 033613 (2016).

\bibitem{Salasnich} Z. Chen, Y. Li, B. A. Malomed, and L. Salasnich,
Spontaneous symmetry breaking of fundamental states, vortices, and dipoles
in two and one-dimensional linearly coupled traps with cubic
self-attraction, Phys. Rev. A \textbf{96}, 033621 (2016).

\bibitem{Yakimenko} A. Oliinyk, A. Yakimenko, and B. Malomed, Tunneling of
persistent currents in coupled ring-shaped Bose--Einstein condensates, J.
Phys. B: At. Mol. Opt. Phys. \textbf{52}, 225301 (2019).

\bibitem{Yakimenko2} A. Oliinyk, I. Yatsuta, B. Malomed, and A. Yakimenko,
Symmetry breaking in interacting ring-shaped superflows of Bose-Einstein
condensates, Symmetry \textbf{11}, 1312 (2019).

\bibitem{Bruno} A. Escriv\`{a}, A. Mu\~{n}oz Mateo, M. Guilleumas, and B.
Juli\'{a}-D\'{\i}az, Tunneling vortex dynamics in linearly coupled
Bose-Hubbard rings, Phys. Rev. A \textbf{100}, 063621 (2019).

\bibitem{Yakimenko3} A. Oliinyk, B. Malomed, and A. Yakimenko, Nonlinear
dynamics of Josephson vortices in merging superfluid rings, Comm. Nonlin.
Sci. Num. Sim. \textbf{83}, 105113 (2020).

\bibitem{Averin} D. V. Averin, Coulomb blockade in superconducting quantum
point contacts, Phys. Rev. Lett. \textbf{82}, 3685-3688 (1999).

\bibitem{saito2006breaking} H. Saito, Y. Kawaguchi, and M. Ueda, \newblock %
Breaking of chiral symmetry and spontaneous rotation in a spinor
Bose-Einstein condensate, \newblock {\em Physical review letters} \textbf{96}%
, 065302 (2006).

\bibitem{Majorana} M. Diez, I. C. Fulga, D. I. Pikulin, M. Wimmer, A. R.
Akhmerov, and C. W. J. Beenakker, Phase-locked magnetoconductance
oscillations as a probe of Majorana edge states, Phys. Rev. B \textbf{87},
125406 (2013).

\bibitem{Kivshar} M. Liu, D. A. Powell, I. V. Shadrivov, M. Lapine, and Y.
S. Kivshar, Spontaneous chiral symmetry breaking in metamaterials, Nature
Commun. \textbf{5}, 4441 (2014)

\bibitem{Weyl} U. Khanna, D. K. Mukherjee, A. Kundu, and S. Rao, Chiral
nodes and oscillations in the Josephson current in Weyl semimetals, Phys.
Rev. B \textbf{93}, 121409 (2016).

\bibitem{Law} C.-Z. Chen, J. J. He, D.-H. Xu, and K. T. Law, Emergent
Josephson current of $N=1$ chiral topological superconductor in quantum
anomalous Hall insulator/superconductor heterostructures, Phys. Rev. B
\textbf{98}, 165439 (2018).

\bibitem{SOC1} Y.-J. Lin, K. Jim\'{e}nez-Garc\'{\i}a, and I. B. Spielman,
Spin-orbit-coupled Bose-Einstein condensates, Nature \textbf{471}, 83-86
(2011).

\bibitem{SOC2} V. Galitski and I. B. Spielman, Spin-orbit coupling in
quantum gases, Nature \textbf{494}, 49 (2013).

\bibitem{SOC3} N. Goldman, G. Juzeli\={u}nas, P. \"{O}hberg, and I. B.
Spielman, Light-induced gauge fields for ultracold atoms, Rep. Prog. Phys.
\textbf{77}, 126401 (2014).

\bibitem{SOC4} H. Zhai, Degenerate quantum gases with spin-orbit coupling: a
review, Rep. Prog. Phys. \textbf{78}, 026001 (2015).

\bibitem{WMLiu} X.-F. Zhang, R.-F. Dong, T. Liu, W. M. Liu, and S.-G. Zhang,
Spin-orbit-coupled Bose-Einstein condensates confined in concentrically
coupled annular traps, Phys. Rev. A \textbf{86}, 063628 (2012).

\bibitem{Rashba} Yu. A. Bychkov and E. I. Rashba, Oscillatory effects and
the magnetic susceptibility of carriers in inversion layers, J. Phys. C
\textbf{17}, 6039 (1984).

\bibitem{Petrov1} D. S. Petrov, Quantum mechanical stabilization of a
collapsing Bose-Bose mixture, Phys. Rev. Lett. \textbf{115}, 155302 (2015).

\bibitem{Fibich} G. Fibich, \textit{The Nonlinear Schr\"{o}dinger Equation:
Singular Solutions and Optical Collapse} (Springer, Heidelberg, 2015).

\bibitem{Sakaguchi_2014} H. Sakaguchi, B. Li, and B.~A. Malomed. \newblock %
Creation of two-dimensional composite solitons in spin-orbit-coupled
self-attractive Bose-Einstein condensates in free space.
\newblock {\em
Phys. Rev. E}, 89, 032920 (2014).

\bibitem{half} B. Ramachandhran, B. Opanchuk, X.-J. Liu, H. Pu, P. D.
Drummond, and H. Hu, Half-quantum vortex state in a spin-orbit-coupled
Bose-Einstein condensate, Phys. Rev. A \textbf{85}, 023606 (2012).

\bibitem{transfer1} M. F. Andersen, C. Ryu, P. Clad\'{e}, V. Natarajan, A.
Vaziri, K. Helmerson, and W. D. Phillips, Quantized rotation of atoms from
photons with orbital angular momentum, Phys. Rev. Lett. \textbf{97}, 170406
(2006).

\bibitem{transfer2} R. Pugatch, M. Shuker, O. Firstenberg, A. Ron, and N.
Davidson, Topological stability of stored optical vortices, Phys. Rev. Lett.
\textbf{98}, 203601 (2007).

\bibitem{transfer3} N. Lo Gullo, S. McEndoo, T. Busch, and M. Paternostro,
Vortex entanglement in Bose-Einstein condensates coupled to Laguerre-Gauss
beams, Phys. Rev. A \textbf{81}, 053625 (2010).

\bibitem{Randy} J. H. V. Nguyen, P. Dyke, D. Luo, B. A. Malomed, and R. G.
Hulet, Collisions of matter-wave solitons, Nature Phys. \textbf{10}, 918-922
(2014).

\bibitem{Sherman} H. Sakaguchi, E. Ya. Sherman, and B. A. Malomed, Vortex
solitons in two-dimensional spin-orbit coupled Bose-Einstein condensates:
Effects of the Rashba-Dresselhaus coupling and the Zeeman splitting, Phys.
Rev. E \textbf{94}, 032202 (2016).

\bibitem{Dressel} G. Dresselhaus, Spin-Orbit coupling effects in zinc blende
structures, Phys. Rev. \textbf{100}, 580-586 (1955).

\bibitem{Manakov} S. V. Manakov, On the theory of two-dimensional stationary
self-focusing of electromagnetic waves, Sov. Phys. JETP \textbf{38}, 248-253
(1974) [Zh. Eksp. Teor. Fiz. \textbf{65}, 505-516 (1973)].

\bibitem{layers} M. Rader, M. Hebenstreit, and R. E. Zillich,\
Multicomponent correlated-basis-function method and its application to
multilayered dipolar Bose gases, Phys. Rev. A \textbf{95}, 033625 (2017).

\bibitem{Arik} A. Gubeskys and B. A. Malomed, Spontaneous soliton symmetry
breaking in two-dimensional coupled Bose-Einstein condensates supported by
optical lattices. Phys. Rev. A \textbf{76}, 043623 (2007).

\bibitem{yang2010nonlinear} J. Yang,
\newblock {\em Nonlinear waves in
integrable and nonintegrable systems}, \newblock SIAM, 2010.


\bibitem{nondiffr0} J. Durnin, Exact solutions for nondiffractive beams. 1.
The scalar theory, J. Opt. Soc. Am. A \textbf{4}, 651-654 (1987).

\bibitem{nondiffr1} D. McGloin and K. Dholakia, Bessel beams: diffraction in
a new light, Contempr. Phys. \textbf{46}, 15-28 (2005).

\bibitem{nondiffr2} A. Ortiz-Ambriz, S. Lopez-Aguayo, Y. V. Kartashov, V. A.
Vysloukh, D. Petrov, H. Garcia-Gracia, J. C. Gutierrez-Vega, and L. Torner,
Generation of arbitrary complex quasi-non-diffracting optical patterns, Opt.
Exp. \textbf{21}, 22221-22231 (2013).

\bibitem{frontiers} R. Zhong, Z. Chen, C. Huang, Z. Luo, H, Tan, B. A.
Malomed, and Y. Li, Self-trapping under the two-dimensional
spin-orbit-coupling and spatially growing repulsive nonlinearity, Front.
Phys. \textbf{13}, 130311 (2018).

\bibitem{embedded} A. R. Champneys, B. A. Malomed, J. Yang, and D. J. Kaup.
``Embedded solitons" : solitary waves in resonance with the linear spectrum,
Physica D \textbf{152-153}, 340-354 (2001).

\bibitem{Romania} H. Sakaguchi, B. Li, E. Ya. Sherman, and B. A. Malomed,
Composite solitons in two-dimensional spin-orbit coupled self-attractive
Bose-Einstein condensates in free space, Romanian Reports in Physics \textbf{%
70}, 502 (2018).

\bibitem{Petrov2} D. S. Petrov and G. E. Astrakharchik, Ultradilute
low-dimensional liquids, Phys. Rev. Lett. \textbf{117}, 100401 (2016).

\bibitem{exp1} I. Ferrier-Barbut, H. Kadau, M. Schmitt, M. Wenzel, and T.
Pfau, Observation of quantum droplets in a strongly dipolar Bose gas, Phys.
Rev. Lett. \textbf{116}, 215301 (2016).

\bibitem{exp2} L. Chomaz, S. Baier, D. Petter, M. J. Mark, F. W\"{a}htler,
L. Santos, F. Ferlaino, Quantum-fluctuation-driven crossover from a dilute
Bose-Einstein condensate to a macrodroplet in a dipolar quantum fluid, Phys.
Rev. X \textbf{6}, 041039 (2016).

\bibitem{exp3} C. R. Cabrera, L. Tanzi, J. Sanz, B. Naylor, P. Thomas, P.
Cheiney, and L. Tarruell, Quantum liquid droplets in a mixture of
Bose-Einstein condensates, Science \textbf{359}, 301-304 (2018).

\bibitem{exp4} G. Semeghini, G. Ferioli, L. Masi, C. Mazzinghi, L. Wolswijk,
F. Minardi, M. Modugno, G. Modugno, M. Inguscio, and M. Fattori, Self-bound
quantum

droplets in atomic mixtures, Phys. Rev. Lett. \textbf{120}, 235301 (2018).

\bibitem{exp5} C. D'Errico, A. Burchiant, M. Prevedelli, L. Salasnich, F.
Ancilotto, M. Modugno, F. Minardi, and C. Fort, Observation of quantum
droplets in a heteronuclear bosonic mixture, Phys. Rev. Research \textbf{1},
033155 (2019).

\bibitem{we} Y. Li, Z. Luo, Y. Liu, Z. Chen, C. Huang, S. Fu, H. Tan, and B.
A. Malomed, Two-dimensional solitons and quantum droplets supported by
competing self- and cross-interactions in spin-orbit-coupled condensates,
New J. Phys. \textbf{19}, 113043 (2017).

\bibitem{OptLett} Y. V. Kartashov, B. A. Malomed, V. V. Konotop, V. E.
Lobanov, and L. Torner, Stabilization of solitons in bulk Kerr media by
dispersive coupling, Opt. Lett. \textbf{40}, 1045-1048 (2015).

\bibitem{Elad} E. Shamriz, Z. Chen, and B. A. Malomed, Stabilization of
one-dimensional Townes solitons by spin-orbit coupling in a dual-core
system, Comm. Nonlin. Sci. Numer. Simul., in press.
\end{thebibliography}
\end{document}